# Structural studies of the sol-gel glasses with copper selenide nanoparticles by SANS technique

V.S. Gurin[1], A.V. Rutkauskas[2], Yu.E. Gorshkova[2,4], S.E. Kichanov[2], A.A. Alexeenko[3], D.P. Kozlenko[2]

[1] Research Institute for Physical Chemical Problems, Minsk, Belarus; gurin@bsu.by

[2] Joint Institute for Nuclear Research, Dubna, Russia

[3] Gomel State Technical University, Gomel, Belarus

[4] Kazan Federal University, Kazan, Russia

**Abstract**

The sol-gel silica glasses doped with copper selenide nanoparticles were studied through small-angle neutron scattering (SANS) and transmission electron microscopy (TEM). SANS intensities demonstrate the complicated dependence on the scattering parameter (transferred momentum within the framework of this technique), however, with no pronounced maxima throughout the full q interval. The multicomponent power-law-exponential model was applied for the approximation and evaluation of the scattering objects sizes at different levels (up to hundreds of nm). Also, fitting to the monoexponential fractal model was elaborated for SANS data. From the results of both simulations we derive the sizes of copper selenide nanoparticles about 60 nm and the noticeable modification of the glass matrix at the scale of the one order more (the long-range effect of particles upon the glass). The fractal dimension of the above scattering objects is varied from 1.5 through 4.2 indicating the formation of complex structure „particle-in-glass" which may not be reduced to a simple model of single insolated particles in solid matrix.

# 1. Introduction

Copper chalcogenides, among many other perspective semiconductors, reveal novel properties combining both quantum confinement effects and the plasmon resonance phenomenon [1-3]. The latter appeared due to elevated free carrier concentration (electrons or holes, depending type of compounds). These compounds possess variable composition and non-stoichiometry and include a number of stable and metastable phases, e.g. in $Cu_{2-x}Se$ for $1 \leq x \leq 2$. The value of $x$ strongly affects the carrier concentration and all their properties. In optics, new absorption bands and IR luminescence were discovered for the chalcogenides and associated with the plasmon resonance. Research of last years has demonstrated very interesting interplay in the nanoscale effects inherent to both metals and semiconductors.

In the present short communication, we investigate the structural features of copper selenide nanoparticles embedded into amorphous (glass-like) silica matrix fabricated through the sol-gel technique [4,5] using small-angle neutron scattering (SANS). This technique allows address the structural data for a wide scale of structural features in materials of different chemical nature (1-1000 nm) with no material destruction [6,7].

# 2. Experimental

The fabrication procedure is based on the sol-gel technique with hydrolysis of tetraethoxysilane resulting in the glassy silica with doping of copper salt precursor with subsequent reduction of copper followed by its selenization. It has been described earlier [4,5]. An important point of this technique is the final step of copper selenization within the silica matrix when a sample is located in the sealed quartz ampoule. This protocol of $Cu_xSe$-doped silica fabrication allow to control both concentration of copper selenide inside and Cu/Se ratio. The latter determines the phase composition of copper selenide nanoparticles and optical features of the semiconductor nanoparticles. The size of particles appears to be in the range of tens of nanometers. They are basically isolated and can form some aggregates at the high doping concentration [6].

The transmission electron microscopy (TEM) study was carried out through the "replica with extraction" procedure. A thin carbon film (10-20 nm) was evaporated onto a freshly etched (diluted HF) surface of a sample followed by the carbon film detaching in water and transfer it to a TEM supporting grid.

SANS measurements were performed on the YUMO spectrometer (IBR-2 high-pulse reactor, Dubna, Russia) [8]. The transferred momentum (q) interval was 0.006 - 0.6 Е$^{-1}$ allowing the considerable wide size range of scattering objects. The data were recorded at the room temperature, they were corrected taking into account the samples transmission and background scattering due to the support, a vanadium ethalone was used. The SasView software [9] was utilized for data processing.

## 3. Results and discussion

Fig. 1 demonstrates the approximated SANS curve for a glass sample with copper selenide nanoparticles. The approximation of the raw SANS data was carried out on the basis of the exponential-power law approach which incorporates both Guinier and Porod dependencies for SANS with smooth matching [10]:

$$I(q) = G_1 exp\left(\frac{-q^2 R_{g1}^2}{3}\right) + B_1 exp\left(\frac{-q^2 R_{g1}^2}{3}\right)\left(\frac{1}{q^*_1}\right)^{P_1} + G_2 exp\left(\frac{-q^2 R_{g2}^2}{3}\right) + B_2 exp\left(\frac{-q^2 R_{g2}^2}{3}\right)\left(\frac{1}{q^*_2}\right)^{P_2} \quad (1),$$

Parameters $G_1$, $G_2$, $B_1$, $B_2$ and exponents $P_1$ и $P_2$ are ajustable in this model, $R_{g1}$ and $R_{g2}$ are giration radia for two structural levels. $q^*_1$ and $q^*_2$ are expressed as:

$$q_1^* = \frac{q}{\left[\text{erf}\left(\frac{k_1 q R_{g1}}{\sqrt{6}}\right)\right]^3} \text{ and } q_2^* = \frac{q}{\left[\text{erf}\left(\frac{k_2 q R_{g2}}{\sqrt{6}}\right)\right]^3} \quad (2)$$

The linear parts were fitted by the formula

$$I = Aq^{-\alpha} + B, \quad (3)$$

where A and B are fitting constants. This evaluation resulted in finding fractal dimensions for different structural levels of the system.

The simulation results according to the model with the formulas (1-2) have shown in Fig. 1 by the curve (no raw measurement data are give here). The structural parameters of scattering objects deduced from this simulation, $Rg_1$ and $Rg_2$, are 253 and 22 nm, respectively. $Rg_2$ value can be assigned to the nanoparticles while $Rg_1$, which is approximately larger on the order of magnitude, may be associated with the areas around them. TEM images (Fig. 2) demonstrate clearly appearance of modifications in the matrix about the particles. However, $Rg_1$ is considerable larger than the observed areas in Fig. 2. In the other words, according to the SANS data, the modifications in the matrix are much wider than the closest cavities in Fig. 2. There is a long-range effect of the particles upon the glassy silica. The strong effect can be supposed from the difference of thermal behavior of $SiO_2$ and $Cu_2Se$ under cooling down from the maximum temperature of the doped glass fabrication, 1200°C. Copper selenide at this temperature exists in the melted state (melting point is 1113°C) and its crystallization may provide an essential change of the particle volume leading to deformation of the glass around. The diameter of copper selenide nanoparticles assuming their sperical shape and the relation with giration radius $R=(5/3)^{1/2}R_g$ can be determined about 57 nm. This value is smaller than the particles on TEM images, but the general consistence may be accepted as sufficient since the SANS model incorporate the wider range sizes for of scattering objects while TEM contrast resolves the maximum contrast ones. A further development of SANS treatment incorporating more structural levels (than two ones here) is expected to result in better agreement of the different measurement techniques.

Fractal properties of the stuctures under study appear to be more complicated, and at least three different exponents may be deduced from the SANS data. The maximum value of α (~4.2) is for the lower q range that corresponds to the larger scattering objects and α~2.2 is obtained for the q range around 0.1 Å$^{-1}$. Likely, these are nanoparticles and their comparatively low fractal dimension can mean the rough particle-matrix interface. The lower α for the intermediate q range and its significant growth further indicate possible variation of the glass structure within the considerable wide region around the particles.

## 4. Conclusion

Thus, the presented SANS study of the silica sol-gel glasses doped with copper selenide nanoparticles has evidenced the formation of the complex scattering objects at the nanometer scale. The good fitting of the experimental data was shown to the model which include different structural levels with Gunier and Porod parts. Fractal properties of the material from these SANS data were also analysed. The $Cu_2Se$ nanoparticles of the average size about 60 nm are principal constituent of these glasses. They modify the closest environment of the silica matrix and the also the long-range effect upon its structure occurs in accordance with the model. The fractaldimension of the particle-matrix system is varied from 1.5 to 4.2 corresponding the complicated ierface and the modified silica matrix.

**Conflict of interest**

The authors declare no conflict of interest.

**Acknowledgements**

The work was performed under partial support from the State Program of Scientific Investigations of Belarus „Material Science, new materials and technologies", Subprogram „Nanostructured materials, nanotechnology, nanotechnik" („Nanostructure").

**Figures**

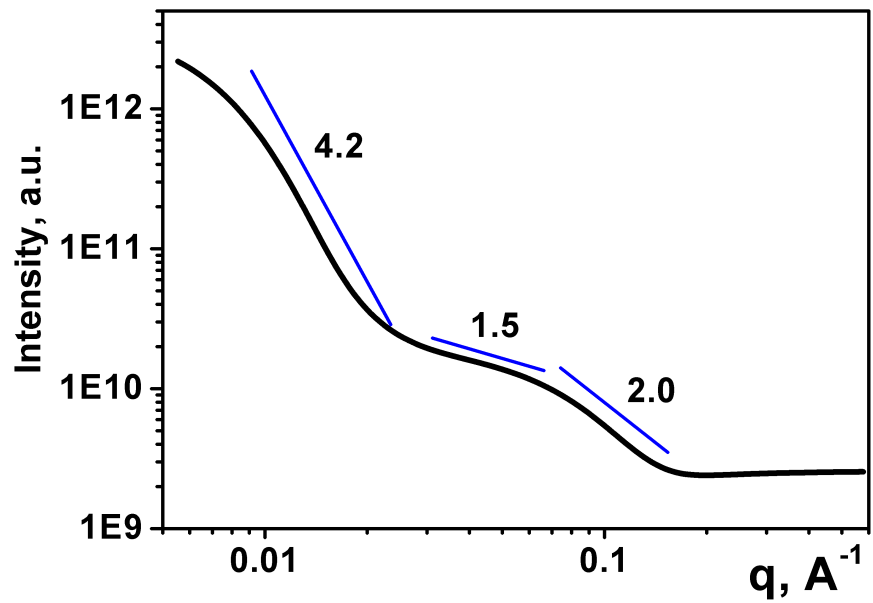

Fig. 1. The data of SANS measurements approximated with the model (1) for a glass with $Cu_2Se$ nanoparticles. The numbers are α exponents for linear fitting with the formula (2) of the corresponding parts of the curve.

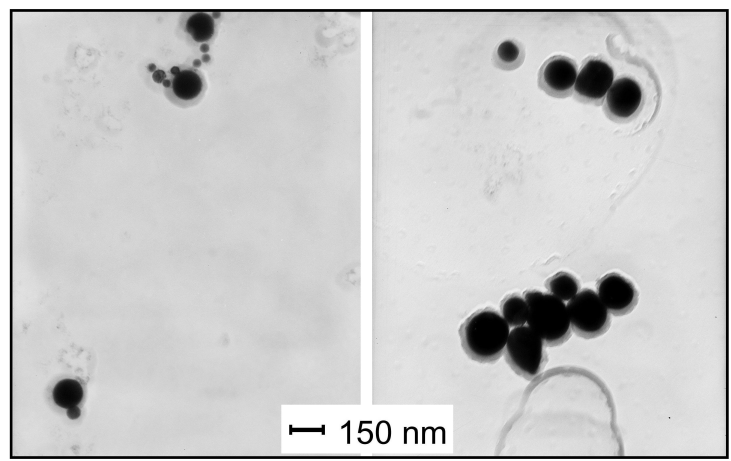

Fig. 2. TEM images of two samples of the glasses with $Cu_2Se$ nanoparticles with different particles concentration. The right sample is close to that analyzed by SANS above.